\documentclass[useAMS,usenatbib,usegraphicx]{mn2e}
\usepackage{psfig}   
\usepackage{graphicx}
\usepackage{amssymb}
\usepackage{subfigure}

\title[Planets around G-dwarfs]{On the frequency of planetary systems around G-dwarfs}

\author[R. J. Parker and S. P. Quanz]{Richard J.~Parker\thanks{E-mail: rparker@phys.ethz.ch} and Sascha P. Quanz \vspace*{0.1cm}\\
   Institute for Astronomy, ETH Z{\"u}rich, Wolfgang-Pauli-Strasse 27, 8093 Z{\"u}rich, Switzerland}

\begin{document}

\date{Accepted for publication in MNRAS}
                             
\pagerange{\pageref{firstpage}--\pageref{lastpage}} \pubyear{2013}

\maketitle

\label{firstpage}

\begin{abstract}
We determine the fraction of G-dwarf stars that could host stable planetary systems based on the observed properties of binaries in the Galactic field, and in various postulated primordial binary populations, which assume that the primordial binary fraction is 
higher than that in the field. We first consider the frequency of Solar System analogues -- planetary systems that form either around a single G-dwarf star, or a binary containing a G-dwarf where the binary separation exceeds 100--300\,au. If the primordial binary fraction 
and period distribution is similar to that in the field, then up to 63\,per cent of G-dwarf systems could potentially host a Solar System analogue. However, if the primordial binary fraction is higher, the fraction of G-dwarf systems that could host a planetary system like our own is lowered to 38\,per cent. 

We extend our analysis to consider the fraction of G-dwarf systems (both single and binary) that can host either circumprimary planets (orbiting the primary star of the binary) or circumbinary planets (orbiting both stars in the binary) for fiducial planetary separations between 1 -- 100\,au. 
Depending on the assumed binary population, in the circumprimary case between 65 and 95\,per cent of systems can host a planet at 1\,au, decreasing to between 20 and 65\,per cent of systems that can host a planet at 100\,au. In the circumbinary case, between 5 and 59\,per cent of 
systems can host a planet at 1\,au, increasing to between 34 and 75\,per cent of systems that can host a planet at 100\,au. 

Our results suggest that the assumed binary fraction is the key parameter in determining the fraction of potentially stable planetary systems in G-dwarf systems and that using the present-day value may lead to significant overestimates if the binary fraction was initially higher.
\end{abstract}


\section{Introduction}
How many stars harbour planetary systems and what fraction of those planets are potentially habitable? In recent years, astronomical observations have made great progress in addressing and partly answering these fundamental questions of astrophysical research. Large-scale, ground-based radial velocity (RV) surveys constrain the occurrence rate of exoplanets around nearby G- and M-dwarf stars for the innermost few astronomical units (au) \citep[e.g.,][]{Mayor11,Bonfils13}, and some first estimates for the mass and period distribution of those planets have been made \citep{Cumming08}. In addition, the \emph{Kepler} space mission was specifically designed to determine the frequency of Earth-sized planets in and near the habitable zones of Sun-like stars \citep{Borucki10} and the first candidates for exoplanets orbiting in their star's habitable zone have been identified \citep[e.g.][]{Borucki12,Borucki13}. 

One important open issue is the question of how the multiplicity of stars impacts the formation and occurrence rate of extrasolar planets. Most RV surveys focus on observing quiet, single stars \citep[e.g.][]{Mayor11}, and for the \emph{Kepler} targets it is unclear how many of the exoplanet candidates potentially orbit binary or multiple star systems. Therefore, the statistical ground on which predictions for planets in multiple systems are made is much weaker. 

Binaries in which a planet orbits one of the stellar components of the system \citep[e.g.][]{Eggenberger04,Raghavan06} have been detected, but the fraction of binary systems that host planets could be much higher than the currently observed value ($\sim 20-30$\,per cent), due to observational incompleteness \citep{Bonavita07}. 
The \emph{Kepler} mission has detected circumbinary planets, where the planet orbits both components of the system. Recent results have shown several planets orbiting binaries with semi-major axes $<$0.25\,au \citep{Doyle11,Orosz12a,Orosz12b,Welsh12}, and \citet{Welsh12} estimate that the frequency of binaries 
with circumbinary Jupiter-mass planets  could be several percent.

The issue of binarity becomes important when one considers the fact that a large fraction of stars in the Galactic field are in binary systems \citep{Duquennoy91,Raghavan10}. The binary fraction (number of binary systems divided by number of single stars plus binary systems) for stars with a similar mass to the Sun (G-dwarfs) is $\sim50$\,per cent \citep{Raghavan10}, but may be lower (30 -- 40\,per cent) for the more numerous M-dwarfs \citep{Fischer92,Lada06,Bergfors10,Janson12}. 
In addition, if some fraction of star formation occurs in relatively dense clustered environments, then interactions between stars can break up binaries with large semi-major axes \citep[e.g.][]{Kroupa95a}.  This implies that the 
primordial binary fraction is probably higher, and recent numerical experiments have suggested a primordial binary fraction of $\sim$75\,per cent \citep{Parker11c,King12a}. 

Hence, if the majority of stars form in binary systems, then the potential effects of binaries on the formation, evolution and stability of planetary systems become important for planetary population synthesis and characterisation. So far, no complete census of the Galactic field is available in order to determine the fraction of binary systems that can host stable planetary systems. Up to now, several authors have focused on 
specific nearby binary systems \citep*[e.g.][]{Jaime12}, or even determined the habitable zones for terrestrial planets in these binaries \citep{Eggl13}. 

In this paper, we adopt a statistical approach with the goal of estimating what fraction of G-type, i.e. solar type, systems -- be they single or binary -- can harbour planetary systems. To do so we make use of the recently updated binary statistics for G-dwarfs in the field \citep{Raghavan10} and we expand the study to consider higher initial binary fractions and different binary semi-major axes distributions. 

The paper is organised as follows. In Section~\ref{bin_props}, we discuss the properties and fraction of binary stars in the field, and in star forming regions. In 
Section~\ref{planet_form} we briefly review the observations of planets, and protoplanetary discs, in binaries. We describe our numerical approach in Section~\ref{method} and present the results of Monte Carlo experiments in Section~\ref{results}. 
We initially focus on the implications of binary properties for Solar System analogues, before discussing the more diverse exoplanetary systems. We present our discussion and conclusions in Section~\ref{conclude}.

\section{Properties of binary systems}
\label{bin_props}

In this section we review the observed properties of binary stars in both the Galactic field and in young star forming regions. We discuss the observed distributions of 
orbital parameters, as well as the overall binary fractions.

\begin{figure*}
  \begin{center}
\setlength{\subfigcapskip}{10pt}
\hspace*{-0.3cm}
\subfigure[Separation distribution of field binaries]{\label{field_bin_sep}\rotatebox{270}{\includegraphics[scale=0.36]{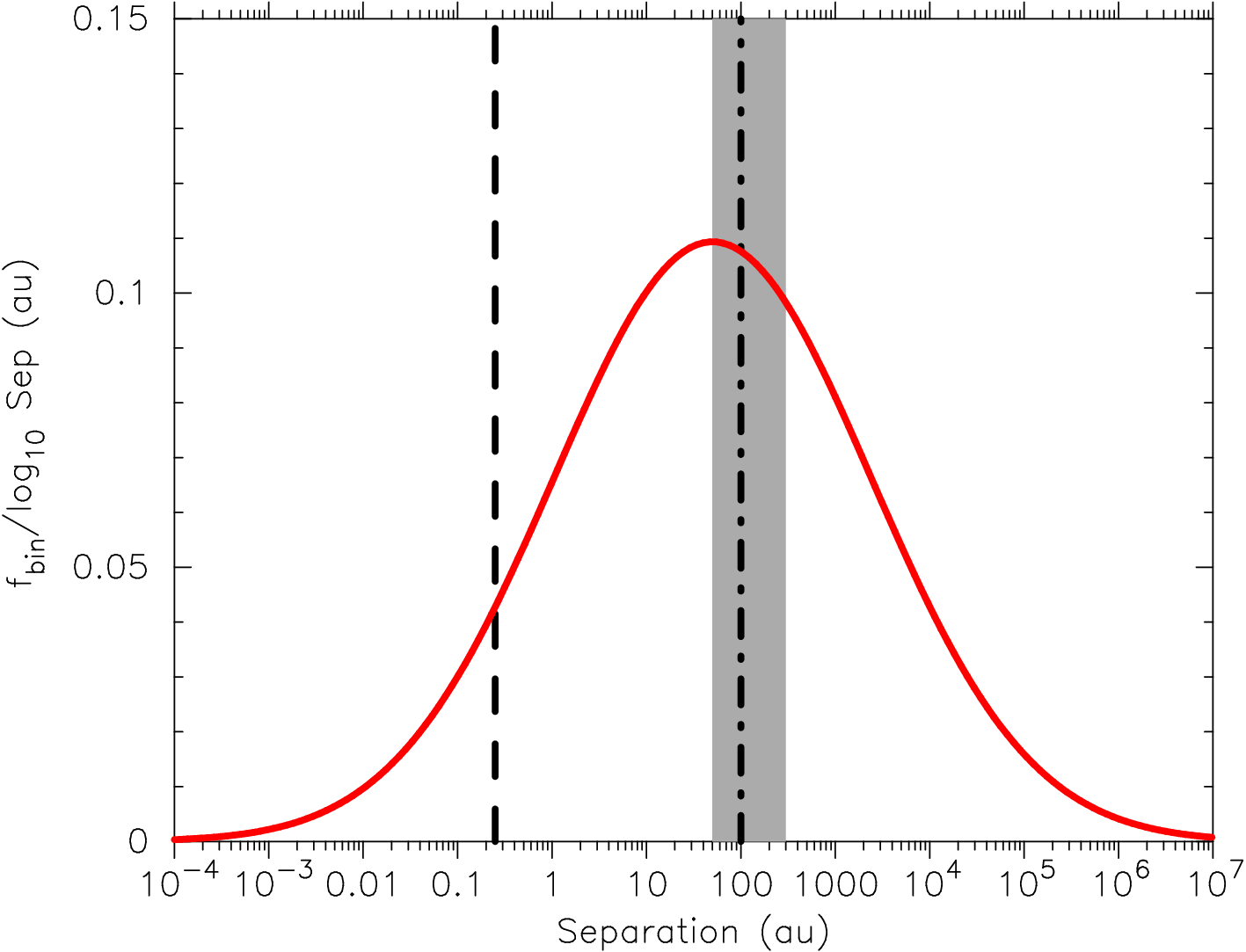}}}  
\hspace*{0.2cm}
\subfigure[Eccentricity--separation distribution of field binaries]{\label{field_bin_esep}\rotatebox{270}{\includegraphics[scale=0.36]{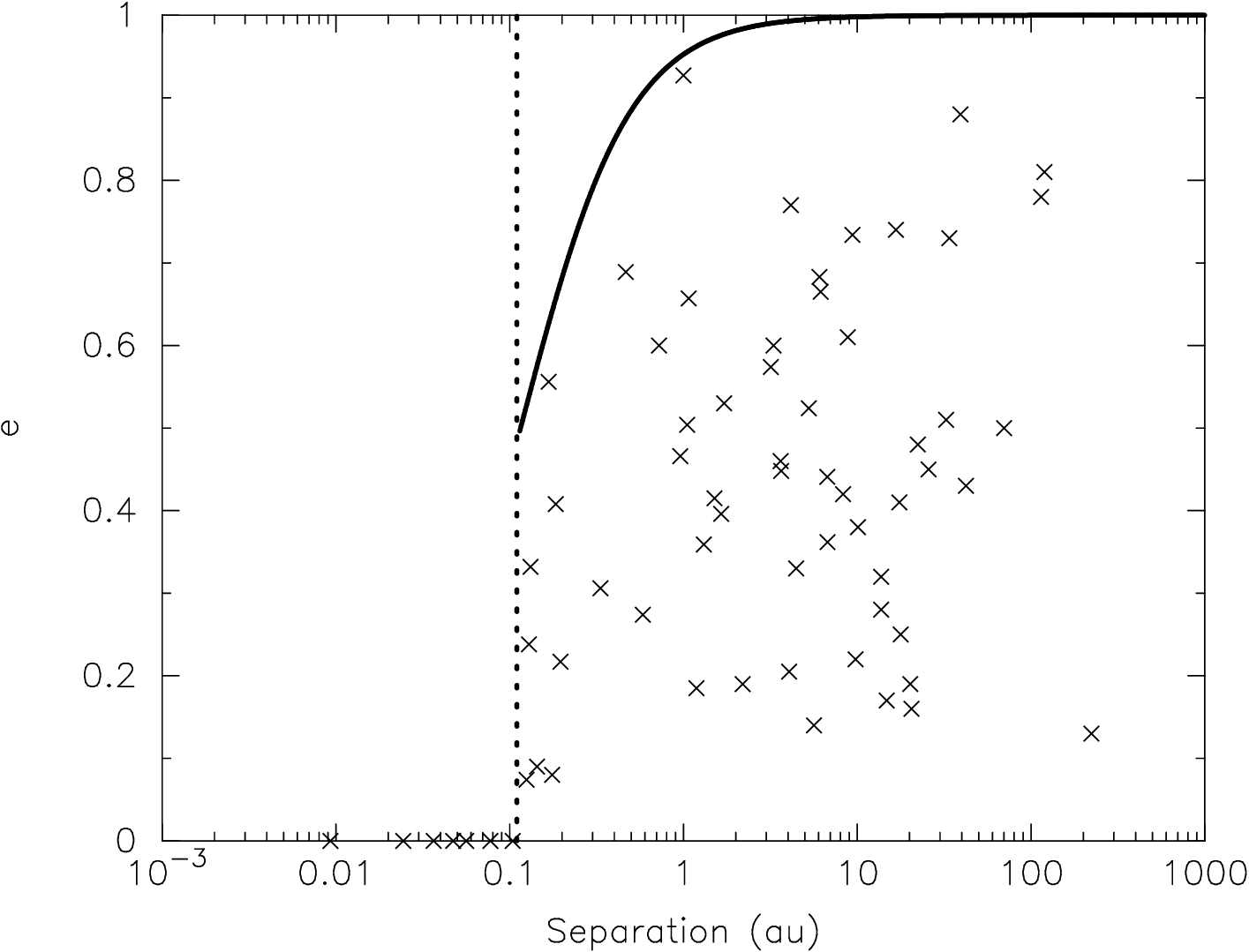}}}
\end{center}
  \caption[bf]{Distribution of orbital parameters of binary stars in the field. The log$_{10}$-normal fit to the separation distribution from \citet{Raghavan10} is shown in panel (a) by the solid red line. The range of minimum binary separations where two planetary systems 
could form independently around both stellar components is shown by the grey shaded region \citep[50 -- 300\,au;][]{Duchene10,Kraus12}, with the `median' value of 100\,au from the literature shown by the dot-dashed line (see Section~\ref{planet_form}). The maximum binary separation for systems hosting circumbinary planets observed by \emph{Kepler}  (0.25\,au) is shown by the dashed line.  The 
relation between orbital eccentricity and separation \citep{Duquennoy91,Raghavan10} is shown in panel (b). The dotted vertical line indicates the separation below which binary orbits are observed to be circular (0.11\,au), and the solid line indicates the transition between eccentricities of zero and a flat distribution.}
  \label{field_binaries}
\end{figure*}

\subsection{The Galactic field}

The binary properties of Solar-like G-type stars (with masses in the range $0.8 \leq m_G/{\rm M_\odot} \leq 1.2$) in the local Solar neighbourhood were comprehensively 
studied in the volume-limited survey by \citet{Duquennoy91}. More recently, \citet{Raghavan10} repeated this work and updated the statistics. The overall binary fraction, $f_{\rm bin}$, of 
G-type stars in the field is 46\,per cent, where
\begin{equation}
f_{\rm bin} = \frac{B}{S + B},
\label{bin_frac}
\end{equation}
and $S$, $B$, are the number of single and binary systems, respectively. Therefore, half of all Sun-like stars reside in binary systems.

Turning to the distributions of orbital parameters, \citet{Duquennoy91} and \citet{Raghavan10} found that the distribution of orbital periods could be fit by a log$_{10}$-normal. Assuming 
an average system mass of 1.5\,M$_\odot$ \citep{Raghavan10}, the semi-major axis (hereafter separation) distribution peaks at 40\,au, but extends from 10$^{-3}$\,au to 10$^6$\,au. The 
 log$_{10}$-normal fit to the separation distribution from \citet{Raghavan10} is shown by the solid line in Fig.~\ref{field_bin_sep}. 

The orbital eccentricities of binary systems vary as a function of the separation (see Fig.~\ref{field_bin_esep}). Very close systems ($a < 0.1$\,au) are thought to have undergone tidal circularisation \citep{Zahn89a,Zahn89b,Mathieu94}, 
whereas wider systems can have an eccentricity of between zero and unity. For wider  ($a > 10$\,au) systems, the eccentricity distribution is consistent with being flat \citep{Raghavan10}.

\subsection{Star forming regions}

Observations of binaries in star forming regions are not as complete as for the nearby field stars. Data on spectroscopic (i.e.\,\,close) binaries are scarce, but observations of visual binaries (with primary masses in the range 0.1 -- 3.0\,M$_\odot$) in nearby 
star forming regions have recently been collated by \citet{King12a,King12b}. The overall binary fraction in nearby star forming regions is generally consistent with the field, although some regions such as Taurus could have a higher fraction. 

The distance to a star forming region governs the separation range for which we observe systems; at small separations 
the two components of a binary will not be resolved and conversely at larger separations the binary becomes indistinguishable against the background cluster members.  
Typically, the separation range probed is $\sim$20 -- 1000\,au, which straddles the peak of the separation distribution of field binaries. However, the study by \citet{King12b} demonstrated that there is an excess of 
binaries with separations in the range 19 -- 100\,au compared to the field. Other authors \citep*[e.g.][]{Connelley08} have noted an apparently flat separation distribution in young star forming regions in the range 100 -- 3000\,au, in agreement 
with {\"O}pik's law \citep{Opik24}. 

Several authors have postulated a very dynamic model of star formation in clusters, in which stars form predominately in binaries ($f_{\rm bin} = 100$\,per cent), which are subsequently destroyed by two-body interactions \citep[e.g.][]{Kroupa95a,Parker11c}. 
\citet{Kroupa95a} and \citet{Kroupa11} formulated a pre-main sequence separation distribution (hereafter K95), in which an excess of binaries with separations $> 200$\,au form and are destroyed in young ($< 10$\,Myr) clusters. However, many of these simulated primordial 
binaries have separations of order the cluster radius \citep{Parker11c} and probably could not form in such dense environments. 

\begin{figure}
\begin{center}
\rotatebox{270}{\includegraphics[scale=0.38]{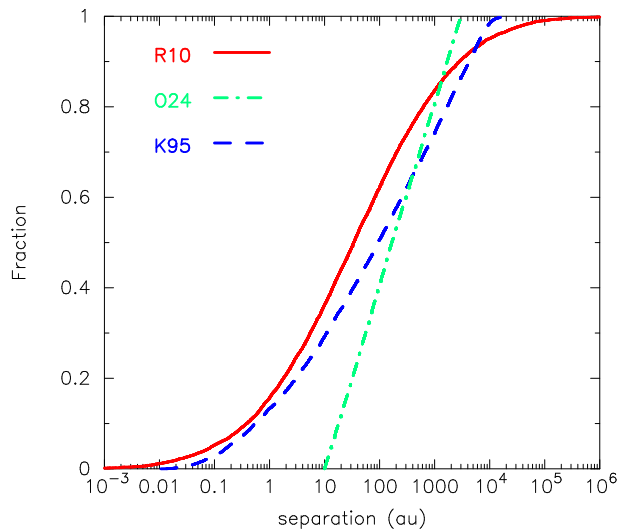}}
\end{center}
\caption[bf]{Three different separation distributions for binary stars. The log-normal fit to G-dwarf binaries in the field from \citet[][R10]{Raghavan10} is shown by the solid red line. The derived pre-main sequence separation distribution from \citet[][K95]{Kroupa95a} 
is shown by the dashed blue line. Finally, a log-uniform distribution between 10\,au and 3000\,au \citep[e.g.][O24]{Opik24,Connelley08} is shown by the dot-dashed green line.}
\label{sma_dists}
\end{figure}

In Fig.~\ref{sma_dists} we show three different separation cumulative distributions. Firstly, the log-normal fit to the \citet{Raghavan10} field G-dwarf binares is shown by the solid red line. Secondly, the pre-main sequence separation distribution derived by 
\citet{Kroupa95a}, which has an excess of wide binaries ($>$200\,au) compared to the field, is shown by the blue dashed line. Finally, a log-uniform separation distribution  \citep[{\"O}pik's law,][]{Opik24} between 10 and 3000\,au \citep[observed in several star forming regions,][]{Connelley08} is shown by the green dot-dashed line.\\

In summary, the fraction of stars that are part of a binary system is $\sim$50\,per cent, but the fraction of stars that \emph{form} in binary systems may be up to 75\,per cent \citep{Parker11c}, with an excess of wide primordial binaries. In the following sections, we will explore the implications of this 
for the formation and stability of planetary systems.

\section{Planets and discs in and around binaries}
\label{planet_form}
Some observational effort has gone into searching for binary companions to known exoplanet hosts \citep[e.g.][]{Raghavan06}. Such planets orbit one of the component stars of the binary in a satellite, or `S-type' 
orbit \citep{Dvorak86}. Estimates of the frequency of binaries hosting planets in `S-type' orbits suggest a 
value of 20 -- 30\,per cent \citep{Raghavan06,Bonavita07}; however, these values are lower-limits and the fraction hosting planets could be similar to single stars \citep{Bonavita07}. Furthermore, recent results from the \emph{Kepler} mission have unearthed 
circumbinary planets \citep[a `P--type' orbit,][]{Dvorak86}; planets orbiting very tight $a_{\rm bin} < 0.25$\,au binary systems \citep{Doyle11,Orosz12a,Orosz12b,Welsh12}.

In order to determine whether a planet could have formed and remained stable in a binary system, we adopt the critical semimajor axis criteria, $a_c$, derived from numerical experiments by \citet{Holman99}. They placed massless particles on coplanar, prograde circular orbits 
within a binary system and obtained a quadratic fit to their numerical results for $a_{c}$ (for an `S-type' orbit): 
\begin{equation}
\begin{array}{lll} a_{cS} & = & [0.464 - 0.38\mu - 0.631e \,\,+ \vspace*{0.1cm} \\  &  & 0.586\mu e + 0.15e^2 - 0.198\mu e^2]a_{\rm bin}  \end{array}
\label{stab_crit_S}
\end{equation}
Here, $a_{\rm bin}$ is the semi-major axis of the binary, $e$ is the eccentricity of the binary, and $\mu = m_s/(m_p + m_s)$, where $m_p$ and $m_s$ are the primary and secondary masses of the binary, respectively. A planet is stable if its semi-major axis is less than $a_{cS}$.

In our analysis, we also consider constraints from observations of discs in binary stars in order to determine the fraction of planetary systems that could form and remain stable in a binary. \citet{Kraus12} analysed the disc frequency of binaries versus single stars in young 
star forming regions, and found that binaries with separations $>50 - 100$\,au are comparable to single stars. Indeed, \citet{Duchene10} notes that binaries with separations $>100 - 300$\,au are indistinguishable from single stars in terms of their protoplantary discs, 
debris discs and fully formed planetary systems. Therefore, a binary with a separation in excess of $\sim$300\,au could potentially harbour two independent planetary systems.

In Fig.~\ref{field_bin_sep}, we show the separation range (50--300\,au) at which binaries become indistinguishable from single stars (in terms of their ability to host planetary systems) by the grey shaded region, and 100\,au is shown by the dot-dashed line.

When considering planets on circumbinary (`P-type') orbits, \citet{Holman99} also derived an expression based on their numerical experiments for the minimum separation a planet could orbit a binary system and remain stable: 
\begin{equation}
\begin{array}{lll} a_{cP} & = & [1.6 +5.1e - 2.22e^2 + 4.12\mu \,\,- \vspace*{0.1cm} \\  &  & 4.27e\mu - 5.09\mu^2 + 4.61e^2\mu^2]a_{\rm bin} \end{array}
\label{stab_crit_P}
\end{equation}
Here, a planet is stable if its semi-major axis is greater than $a_{cP}$. 

The circumbinary planets detected by \emph{Kepler} orbit binaries with semi-major axes $a < 0.25$\,au. However, circumbinary discs have been observed around binaries with much larger separations \citep{Monin07}.  Specific examples include 
SR24N, where $a \sim 32$\,au \citep{Andrews05};  GG\,Tau\,A, where $a \sim 60$\,au \citep{Kohler11}; and UY\,Aur, where $a \sim 190$\,au \citep{Close98}. Whilst only a handful of examples, these systems demonstrate that circumbinary planet formation may occur 
around very wide (10s of au) binary systems.

In our analysis of planets orbiting on such `P-type' orbits, we will exclude the possibility that a system might theoretically contain circumbinary \emph{and} circumprimary planets. 

Equations~\ref{stab_crit_S}~and~\ref{stab_crit_P} are only valid for binaries with eccentricities $e < 0.7 - 0.8$, and for $0.1 \leq \mu \leq 0.9$ \citep{Holman99}. In Section~\ref{results} we discuss the frequency of sampled binaries in our experiment that lie outside this range. 

Note that we are using the terms separation and semi-major axis interchangeably; most observations of binary stars only determine the instantaneous separation, which is related \emph{on average} to the semi-major axis \citep {Duquennoy91} by
\begin{equation}
{\rm log}\,a" = {\rm log}\,\rho" + 0.13,
\end{equation}
where $a"$ and $\rho"$ are the semi-major axis and separation (both in arcseconds). Therefore, when determining whether a binary could host a stable planet according to $a_{cS}$ or $a_{cP}$, care must be taken to consider the full range of binary semi-major axes that may be possible, based on the separation.

The stability criteria derived by \citet{Holman99} assumes that the planets have already formed. The study by \citet{Pichardo05} discusses the truncation of circumstellar discs as a function of the host binary parameters, and \citet{Thebault06} discuss the regime in which binary components affect 
accretion of planetesimals. Indeed, \citet{Duchene10} finds that planet formation in $<100$\,au binaries is affected by much shorter clearing timescales for the protoplanetary discs, perhaps preventing the slow build-up 
of terrestrial planets. In our subsequent analysis, we do not comment on the mass or composition of the hypothetical planets, and assume they have been allowed to form without external perturbations.

Finally, we note that in reality, planets may exist that do not fulfill the stability criteria in Equations~\ref{stab_crit_S}~and~\ref{stab_crit_P}. As an example, the $\nu$\,Oct system is a binary with a semi-major axis $a_{\rm bin} = 2.55$\,au which appears to host a planet with a (relatively) high semi-major axis 
\citep[$a_{\rm planet} = 0.45a_{\rm bin}$,][]{Ramm09}. For the system parameters, a massless particle can only be stable according to Equation~\ref{stab_crit_S} if $a_{\rm planet} < 0.25a_{\rm bin}$, although stability may be possible if the orbit is retrograde \citep{Eberle10,Gozdziewski13}.

\section{Method}
\label{method}

In order to create our field population of single and binary systems, we perform the following Monte Carlo experiment. We choose a random number between 0 and 1 and if it is lower than our chosen $f_{\rm bin}$ we make a binary system. We draw primary masses from a composite IMF, 
consisting of the \citet{Chabrier05} log-normal for $m < 1$\,M$_\odot$, and the \citet{Salpeter55} power-law slope for higher-mass stars \citep[see also][]{Bastian10}:
\begin{equation}
\begin{array}{lll} 
\xi ({\rm log}\,m) & = & 0.093\,{\rm exp}\left \{-\frac{{\left({\rm log}m - \overline{{\rm log}m}\right)}^2}{2\sigma^2_{{\rm  log}m}}\right \}, {\,\,} m \leq 1\,{\rm M}_\odot  \vspace*{0.1cm}\\  & = & 0.041\,m^{-1.35}, \hspace*{2.3cm} m > 1\,{\rm M}_\odot. 
\end{array}
\end{equation}
Here, $\overline{{\rm log}\,m} = 0.2$ is the mean stellar mass and $\sigma_{{\rm  log}\,m} = 0.55$ is the variance \citep{Chabrier05}. 
As we are limiting our study to G-dwarf primaries, we re-select the mass if falls outside of the range $0.8 \leq m_G/{\rm M_\odot} \leq 1.2$.  The mass of the secondary component is drawn from a flat mass ratio distribution, in accordance with observations of binaries in the field \citep{Metchev09,Reggiani11a}. 

For the distribution of the semi-major axes, or equivalently, of the periods of our binary systems, we pick one of the three different distributions presented in Fig.~\ref{sma_dists}:
\begin{itemize}
\item  The log-normal fit to the data in the field by \citet{Raghavan10}:
\begin{equation}
\left({\rm log_{10}}P\right)  \propto {\rm exp}\left \{ \frac{-{({\rm log_{10}}P - \overline{{\rm log_{10}}P})}^2}{2\sigma^2_{{\rm  log_{10}}P}}\right \},
\end{equation}
where $\overline{{\rm log_{10}}P} = 5.03$, $\sigma_{{\rm log_{10}}P} = 2.28$ and $P$ is  in days. 
\item The initial pre-main sequence period function derived by \citet{Kroupa95a}: 
\begin{equation}
f\left({\rm log_{10}}P\right) = \eta\frac{{\rm log_{10}}P - {\rm log_{10}}P_{\rm min}}{\delta + \left({\rm log_{10}}P - {\rm log_{10}}P_{\rm min}\right)^2},
\end{equation}
where ${\rm log_{10}}P_{\rm min}$ is the logarithm of the minimum period in days. We adopt ${\rm log_{10}}P_{\rm min} = 0$; and $\eta = 3.5$ and $\delta = 100$ are the numerical 
constants adopted by \citet{Kroupa95a} and \citet{Kroupa11} to fit the observed pre-main sequence distributions.
\item A log-uniform distribution, first postulated by \citet{Opik24} and observed (in the range 10 -- 3000\,au) in some star forming regions \citep{Connelley08}.
\end{itemize}

We convert the periods to semi-major axes using the masses of the binary components, which, for the orbital periods derived by \citet{Raghavan10}, results in the distribution shown by the solid red line in Fig.~\ref{field_bin_sep}.

We also assign an orbital eccentricity to each binary based on the distribution observed for field G-dwarf binaries. Systems with periods less than 12 days ($\sim 0.11$\,au) are on circular orbits ($e = 0$), whereas systems with longer periods have a flat eccentricity 
distribution \citep{Raghavan10}. We first draw eccentricities from a flat distribution, and circularise systems with $P < 12$\,days. If a system has $P > 12$\,days, but the chosen eccentricity exceeds the following period-dependent value (indicated by the solid line in Fig.~\ref{field_bin_esep}):
\begin{equation}
e_{\rm tid} = \frac{1}{2}\left[1 + {\rm tanh}\left(1.6\,{\rm log_{10}}P - 1.7\right)\right],
\end{equation}
we reselect the eccentricity. Note that this differs slightly from the formula adopted in our previous work \citep[e.g.][]{Parker12a}, which was based on the older \citet{Duquennoy91} data. In star forming regions the eccentricity distributions are not well constrained, 
and we adopt the field distribution for our other assumed period/separation distributions.

We repeat this process until we have a total of 1000 \emph{stars} for each realisation. We now have a `population' of systems (both binary and single) for which we determine the fraction that could harbour a `planetary system', which is defined in different ways. This experiment is repeated 10 times, yielding 10 independent `populations', in order to quantify the stochasticity of this approach and to obtain an average value and an associated variance. 

\section{Results}
\label{results}

In this section we first discuss the frequency of stable Solar System analogues based on the Galactic binary population (Section~\ref{sol_results}), before considering extrasolar systems in general (Section~\ref{exo_results}). We define a Solar system analogue as either a single G-dwarf; or alternatively a binary which contains at least one G-dwarf, and has a separation greater than 300\,au (which would allow our planetary system to exist in a stable configuration). 

\subsection{The Solar System}
\label{sol_results}

\begin{figure}
\begin{center}
\rotatebox{270}{\includegraphics[scale=0.35]{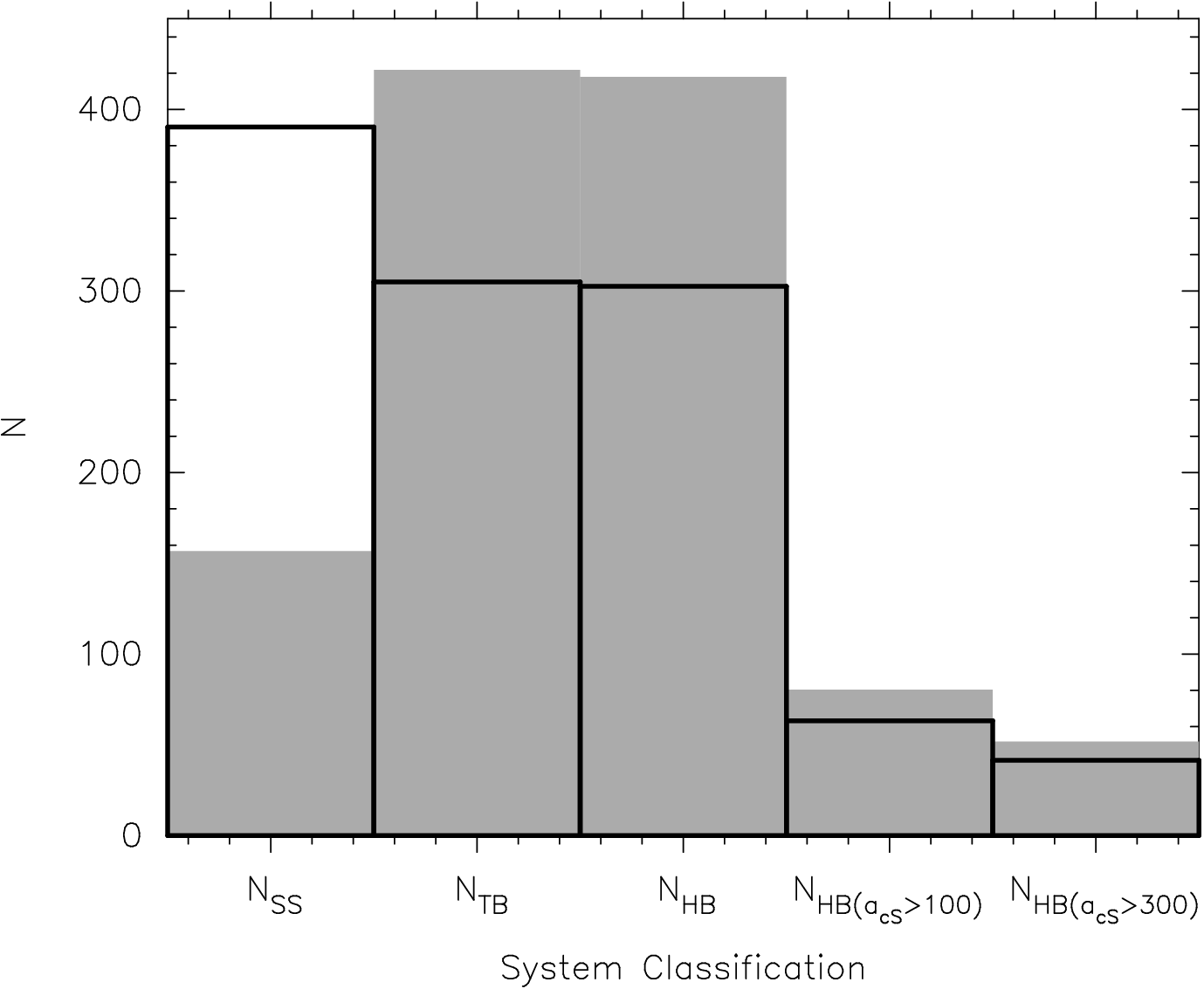}}
\end{center}
\caption[bf]{The distribution of systems for a field-like binary fraction ($f_{\rm bin} = 0.46$) and the \citet{Raghavan10} separation distribution (open histogram), and assuming the same separation distribution but a binary fraction 
of $f_{\rm bin} = 0.75$ (the shaded histogram). The number of single stars ($N_{\rm SS}$) dominates for $f_{\rm bin} = 0.46$, whereas the binaries ($N_{\rm TB}$) dominate for a higher binary fraction. The number of systems for which we can apply the \citet{Holman99} stability criterium is given by $N_{\rm HB}$. Interestingly, the number of binaries that could potentially 
host solar system analogues ($N_{{\rm HB}(a_{cS} >100)}$ and $N_{{\rm HB}(a_{cS} >300)}$) are similar for both populations (see text and Table~\ref{solar_results} for full details and uncertainties).}
\label{sol_hist}
\end{figure}

\begin{table*}
\caption[bf]{Numbers of \emph{systems} from Monte Carlo sampling of 1000 stars with the \citet{Raghavan10} period distribution observed in the field, but with two different binary fractions. From left to right, the binary fraction ($f_{\rm bin}$), the number of single stars with this binary fraction, 
$N_{\rm SS}$, the total number of binary stars, $N_{\rm TB}$, the number of binaries that have a periastron distance $r_{\rm peri}$ exceeding the \citet{Holman99} stability criteria, $N_{\rm HB}$. 
Next, we show the number of these planet-hosting binaries with $a_{cS} > 100$\,au, $N_{{\rm HB}(a_{cS} >100)}$, and the number of host binaries with $a_{cS} > 300$\,au, $N_{{\rm HB}(a_{cS} >300)}$. Finally, we show the 
number of host binaries with $a_{cS} > 100$\,au and $a_{cS} > 300$\,au, that also have a G-dwarf secondary ($N_{{\rm HB}(a_{cS} >100) , Gs}$ and $N_{{\rm HB}(a_{cS} >300) , Gs}$, respectively).}
\begin{center}
\begin{tabular}{|c|c|c|c|c|c|c|c|}
\hline 
$f_{\rm bin}$  & $N_{\rm SS}$ &  $N_{\rm TB}$ & $N_{\rm HB}$ & $N_{{\rm HB}(a_{cS} >100)}$ & $N_{{\rm HB}(a_{cS} >300)}$ & $N_{{\rm HB}(a_{cS} >100) , Gs}$ & $N_{{\rm HB}(a_{cS} >300) , Gs}$\\
\hline
0.46 & 390 $\pm$ 15 & 305 $\pm$ 7 & 303 $\pm$ 7 & 63 $\pm$ 6 & 42 $\pm$ 4 & 10 $\pm$ 1 &  7 $\pm$ 2\\
\hline
0.75 & 157 $\pm$ 12 & 422 $\pm$ 6 & 418  $\pm$ 6 & 80 $\pm$ 7 & 52 $\pm$ 6 & 15 $\pm$ 3 & 10 $\pm$ 4\\
\hline
\end{tabular}
\end{center}
\label{solar_results}
\end{table*}

From our Monte Carlo experiments, we determine the fraction of G-type stars that could host `Solar System analogues', taking into account the binary properties in the field \citep{Raghavan10}. Estimating this fraction is based 
on the following criteria:

Firstly, as the binary fraction of G-type stars is 0.46, a significant number of these stars are single and have no constraints on the planetary systems they can host. We refer to 
the number of single G-type stars as `$N_{\rm SS}$' in Table~\ref{solar_results} and Fig.~\ref{sol_hist}.  

The remaining systems are then binaries and refer to the \emph{total} number of binaries as `$N_{\rm TB}$'. We then apply the \citet{Holman99} critical semi-major axis criterium (see equation 2) to the binary system. If the periastron distance (defined in the usual way as $r_{\rm peri} = a(1 - e)$) exceeds $a_{cS}$, then the binary system can potentially host a planetary system (in an S--type orbit) and we apply Eq.~\ref{stab_crit_S} for the chosen planetary separation. In certain cases, such as high ($>0.9$) binary eccentricity, we cannot apply Eq.~\ref{stab_crit_S} and we must remove the system from our analysis (this turns out to be only a handful of systems -- the difference between columns 3 and 4 in Table~\ref{solar_results}). The number of binary systems that we can apply Eq.~\ref{stab_crit_S} to is designated `$N_{\rm HB}$'. 

Of the $N_{\rm HB}$ binaries, we apply the constraints discussed in Section~\ref{planet_form}. We consider 50 au as the ``edge'' of our Solar System \citep{Allen01} and determine the number of binaries where the critical semi-major axis for planetary stability is $a_{cS} > 100$\,au, which we label  `$N_{{\rm HB}(a_{cS} >100)}$'; and the number of binaries where $a_{cS} > 300$\,au (`$N_{{\rm HB}(a_{cS} >300)}$').
This is the separation regime in which a binary system could host two individual protoplanetary 
discs which have not been truncated, i.e. they can be treated as single two stars \citep{Duchene10}. As we are restricting our definition of a Solar analogue to a system with a G-type primary star, the binaries in which we treat the components as two individual stars 
must have a G-type secondary in order for us to count the system as a pair of Solar Systems. For  $a_{cS} > 100$\,au and $a_{cS} > 300$\,au  the number of binaries that have a G-dwarf secondary is `$N_{{\rm HB}(a_{cS} >100) , Gs}$' and `$N_{{\rm HB}(a_{cS} >300), Gs}$', respectively. 

The results for the binary properties in the Galactic field are shown in Table~\ref{solar_results} and Fig.~\ref{sol_hist}. The open histogram in Fig.~\ref{sol_hist} shows the results for a G-dwarf binary fraction of 0.46. Sampling 1000 stars, we obtain 390($\pm 15$) single stars and 305($\pm 7$) binary systems, 
a total of 695 systems. We are able to apply the \citet{Holman99} criteria for 303 of these binaries. 
The number of binaries that have $a_{cS} >$100\,au and $a_{cS} >$300\,au is $N_{{\rm HB}(a_{cS} >100)} = 63 \pm 6$ and $N_{{\rm HB}(a_{cS} >300)}= 42 \pm 4$, respectively. As mentioned above, a proportion of these latter systems have a G-dwarf secondary, so in principle these binaries could host two Solar System analogues (the final two columns in Table~\ref{solar_results}).

To summarise, the number of G-type \emph{systems} that could host a Solar System analogue (conservatively assuming disc truncation in binaries with $a_{cS} < 300$\,au) is  $N_{\rm SS} + N_{{\rm HB}(a_{cS} >300)} + N_{{\rm HB}(a_{cS} >300), Gs} = 439 \pm 16$, which translates into 63\,per cent of our sample 
of 695 systems\footnote{Note that if we relax this constraint and allow Solar System analogues in binaries with $a_{cS} > 100$\,au, then 67\,per cent of systems can host stable Solar Systems.}. 

However, the primordial binary fraction is likely to have been higher than the value currently observed in the field \citep[e.g.][]{Kroupa95a,Goodwin05a,Kaczmarek11,Parker11c}. If we assume the primordial binary fraction was $\sim$0.75 \citep{Parker11c,King12a}, then the distribution of systems (shown by the grey shaded histogram in Fig.~\ref{sol_hist}) is markedly different (see also the second row of Table~\ref{solar_results}). We now have only 
$N_{\rm SS} = 157(\pm 12)$ single stars, $N_{\rm TB} = 422(\pm 6)$ binary systems (a total of 579 \emph{systems}),  of which $N_{\rm HB} = 418(\pm 6)$ binaries could host stable planets. Again, assuming that a Solar System analogue could form without perturbations from the secondary star if 
$a_{cS} > 300$\,au, we obtain the number of systems that could host a Solar System analogue as $N_{\rm SS} + N_{{\rm HB}(a_{cS} >300)} + N_{{\rm HB}(a_{cS} >300), Gs} = 219 \pm 14$, which is only 38\,per cent of the total number of systems.

\subsection{Extrasolar systems}
\label{exo_results}

We now expand our results to the more general case and consider planets with a range of semi-major axes, and the proportion that can be stable, both in binary systems (S--type orbits) and around binary systems (P--type orbits). 

Our goal here is to evaluate the impact of the assumed stellar binary fraction and separation distribution on the frequency of planets that could exist in either circumprimary or circumbinary orbits. We 
do not synthesise the results together to derive a global frequency of stellar systems that could host planets for two reasons. Firstly, very wide ($>300$\,au) binaries could potentially host two individual planetary systems, but we would require knowldege of 
the architecture of the planetary system, which may not be universal, to determine whether the planets are stable. Secondly, some moderately wide ($\sim 10 - 100$\,au) binaries host circumbinary discs; we would therefore need some criterium to decide whether e.g.\,\,a 30\,au binary will form circumprimary, circumbinary, or both types of, planetary systems in order to compute the global frequency. 

We consider 5 different planetary semi-major axes; 1\,au (an Earth-like orbit), 5\,au (Jupiter-like), 30\,au (Neptune-like), 
50\,au and 100\,au. The final two values are chosen as potential separations for planets that may form via gravitational instability, rather than core accretion -- and hence are able to form further from their host star.

\subsubsection{Circumprimary (S--type) orbits}

Firstly, as in Section~\ref{sol_results}, we draw 1000 stars with a binary fraction of 0.46; this translates into 695 \emph{systems}. Of the systems that are binary, we determine whether we can apply the \citet{Holman99} criteria (we discard the very few binary systems for which $r_{\rm peri} < a_{cS}$). 
We then determine the fraction of \emph{systems}\footnote{Here we do not consider the fraction of \emph{stars} that could form a stable planetary system because this would require an assumption about the frequency of binaries where the secondary star hosts a planetary systems, and the architecture of the planetary systems.} that could host a planet, $f_{\rm stable, S}$, at the chosen planetary separation, $a_p$ distance thus:
\begin{equation}
f_{\rm stable, S} = \frac{N_{a_{cS} > a_p}+N_{\rm SS}}{N_{\rm TB} + N_{\rm SS}},
\label{stable_S}
\end{equation}
where $N_{a_{cS} > a_p}$ is the number of binaries with a critical semi-major axis (Eq.~\ref{stab_crit_S}) greater than the chosen planetary separation. Here we assume that planets form around every single star.

\begin{figure*}
  \begin{center}
\setlength{\subfigcapskip}{10pt}
\hspace*{-0.3cm}
\subfigure[Circumprimary (S--type) orbits]{\label{stable_fracs_circumprime}\rotatebox{270}{\includegraphics[scale=0.36]{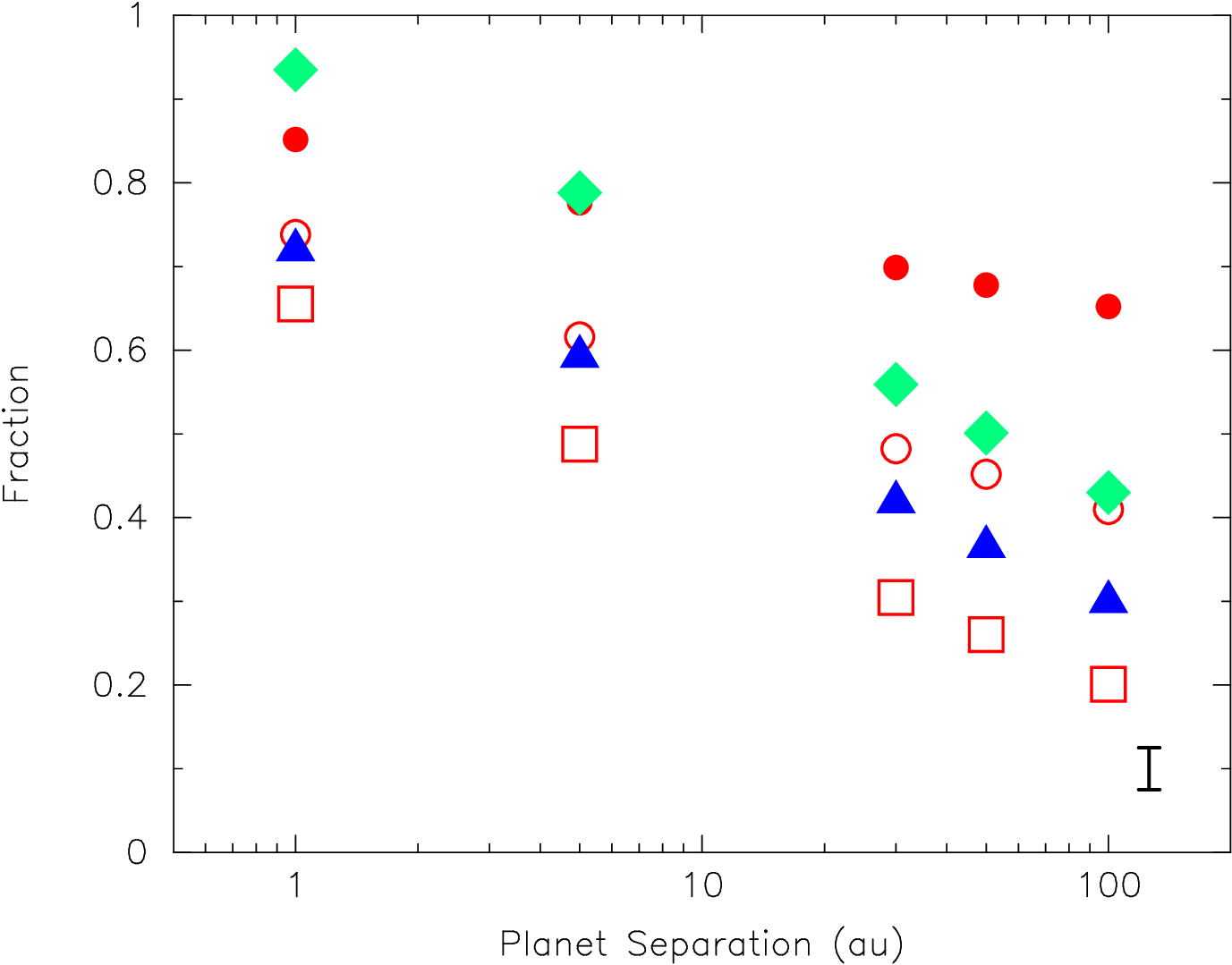}}}  
\hspace*{0.5cm}
\subfigure[Circumbinary (P--type) orbits]{\label{stable_fracs_circumbin}\rotatebox{270}{\includegraphics[scale=0.36]{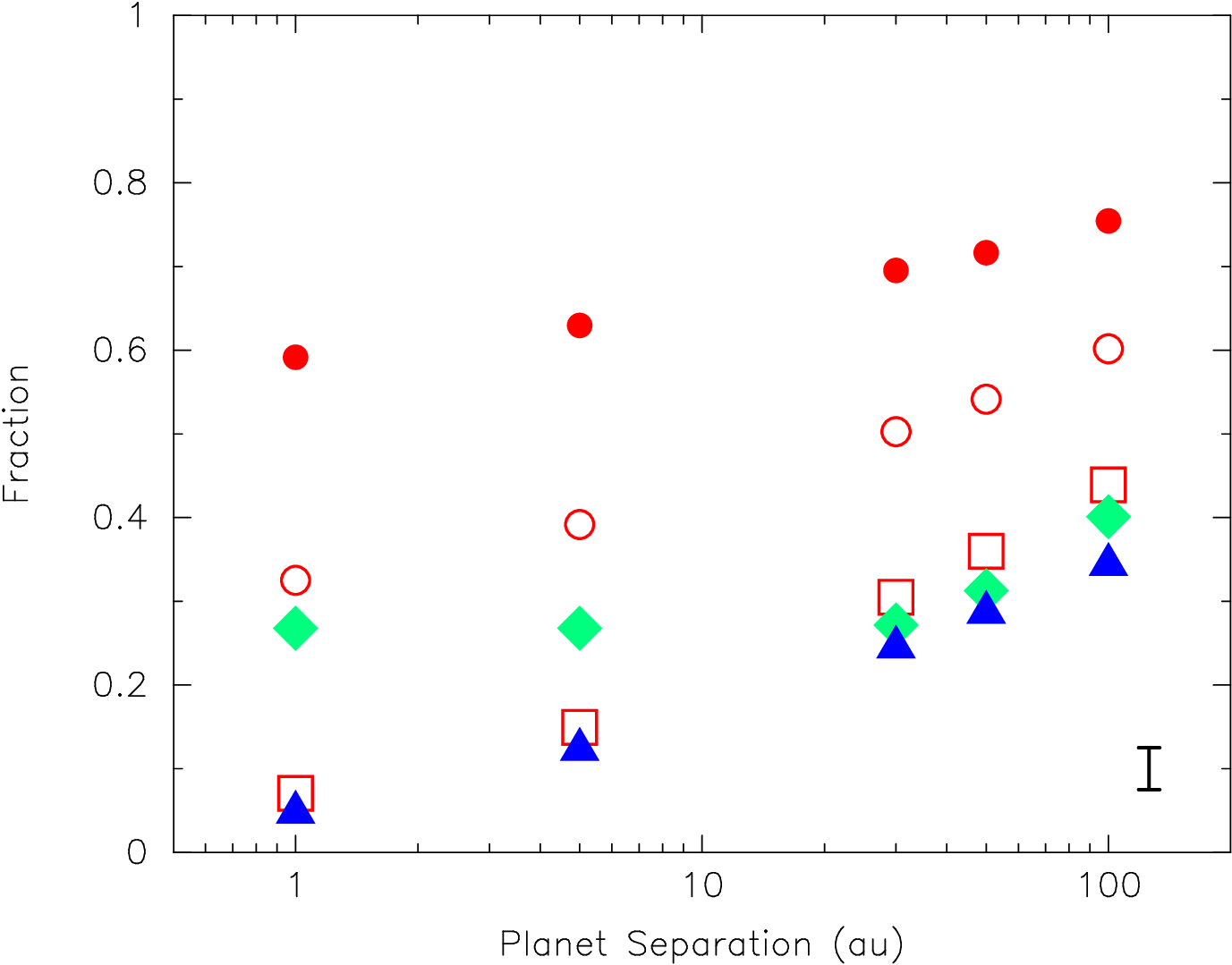}}}
\end{center}
  \caption[bf]{The fraction of \emph{systems} (binaries and single stars) that could host a stable planet at the given separation on either \emph{circumprimary} (S--type) orbits (panel a) or \emph{circumbinary} (P--type) orbits (panel b). 
The filled red circles are for systems with a field-like binary fraction ($f_{\rm bin} = 0.46$) and periods/eccentricities drawn from the \citet{Raghavan10} distributions, whereas 
the open red circles and open red squares show the fraction of systems that could host a stable planet for the same orbital parameters but binary fractions of  $f_{\rm bin} = 0.75$ and  $f_{\rm bin} = 1$, respectively. The results for different 
binary period distributions are also shown. The blue triangles are for binaries with periods drawn from the postulated pre-main sequence distribution in \citet{Kroupa95a} and an overall fraction $f_{\rm bin} = 1$. The green diamonds are 
for binaries with separations drawn from a log-uniform distribution \citep{Opik24,Connelley08} and a binary fraction $f_{\rm bin} = 0.75$. A representative variance from 10 simulations is shown in the bottom right corner of each panel. }
  \label{stable_fracs}
\end{figure*}

In Fig.~\ref{stable_fracs_circumprime} we show the fraction of systems which could host planetary systems, according to Eq.~\ref{stable_S}. We show the fraction for stars with the field statistics ($f_{\rm bin} = 0.46$) by the red solid points. The majority of binaries in the field have a critical semi-major axis much larger than 1\,au, 
and the fraction of systems that can host a planet is high. 85\,per cent of binaries can host a planet with a semi-major axis of 1\,au, which decreases with semi-major axis (70\,per cent of systems can host a planet at 30\,au and 65\,per cent of systems can host a planet at 100\,au). 

Assuming a higher primordial binary fraction than observed in the field ($f_{\rm bin} = 0.75$ versus $f_{\rm bin} = 0.46$) decreases the fraction of systems that can host planets. We show these fractions by the red open circles in Fig.~\ref{stable_fracs_circumprime}; 74\,per cent of systems can host a 
planet at 1\,au, 48\,per cent can host a planet at 30\,au and 41\,per cent of systems can host a planet orbiting at 100\,au. If we assume an even higher binary fraction ($f_{\rm bin} = 1$) with the \citet{Raghavan10} period distribution, the fraction of systems that can host stable planets drops even further (the red open squares in Fig.~\ref{stable_fracs_circumprime}). 

There is a slight dependence on the results when assuming a different binary period distribution. The \citet{Kroupa95a} period distribution for pre-main sequence binaries assumes a binary fraction of unity, and the shape of the distribution is field-like for close binaries, with an excess of binaries with separations $> 100$\,au. The fraction of systems that can host stable planets at 1, 5, 30, 50 and 100\,au assuming $f_{\rm bin} = 1.0$ and the \citet{Kroupa95a} period distribution is shown by the blue triangles in Fig.~\ref{stable_fracs_circumprime}. The fractions of systems that can host stable planets at various separations are similar to the results for the \citet{Raghavan10} period distribution and a lower binary fraction of $f_{\rm bin} = 0.75$ (the red open circles). The reason for the fractions of stable systems being more similar to the \citet{Raghavan10} distribution for $f_{\rm bin} = 0.75$ rather than $f_{\rm bin} = 1.0$ is because the Kroupa distribution contains more binaries with wide ($>$ 100\,au) separations, which can host stable planets at our chosen fiducial separations.

Similarly, there is a dependence on the assumed binary period distribution when we compare a log-uniform distribution (the green diamonds in Fig~\ref{stable_fracs_circumprime}) to a log-normal period distribution with the same binary fraction ($f_{\rm bin} = 0.75$, the open red circles). Here, more planets are stable in binaries with separations drawn from the log-uniform 
distribution; however, we note that this distribution is truncated by observational incompleteness (it only spans the range 10 -- 3000\,au compared to $10^{-3} - 10^6$\,au for binaries in the field) and it is not clear how far it can be extrapolated. 

In summary, depending on the assumed binary population, the fraction of systems that can host a planet on a circumprimary S--type orbit at 1\,au ranges from $\sim$65\,per cent to 95\,per cent. This fraction decreases almost linearly in log-space as a function of planet semi-major axis, so the fraction of systems that can host a planet 
at 100\,au drops to between 20\,per cent and 65\,per cent.

\subsubsection{Circumbinary (P--type) orbits}

We now consider planets that could form on P--type orbits, i.e.\,\,in a disc surrounding both stellar components in the binary. We use the same 695 systems in which we calculated the fraction of planets that could be on stable circumprimary (S--type) orbits and first determine whether the apastron distance ($r_{\rm ap} = a(1 + e)$) 
is less than $a_{cP}$ so that we can apply Eq.~\ref{stab_crit_P} to the system.  

The fraction of systems that could host a planet at the given separation is calculated using:
\begin{equation}
f_{\rm stable, P} = \frac{N_{a_{cP} < a_p}+N_{\rm SS}}{N_{\rm TB} + N_{\rm SS}},
\label{stable_P}
\end{equation}
where $N_{a_{cP} < a_p}$ is the number of binaries with with a critical semi-major axis less than the chosen planetary separation. $N_{\rm TB}$ and $N_{\rm SS}$ are again the total number of binaries and single stars, which vary depending on the assumed binary fraction, $f_{\rm bin}$.

In Fig.~\ref{stable_fracs_circumbin} we show the value of $f_{\rm stable, P}$ for the same binary populations in Fig.~\ref{stable_fracs_circumprime}. As we would expect, the results are qualitatively opposite to the circumprimary S--type case. All of the assumed binary period distributions have median values between 20 -- 100\,au, and the fraction of stable systems is lowest for planet separations of 1\,au, and highest for planets at 100\,au. Assuming the observed field binary properties ($f_{\rm bin} = 0.46$ and a \citet{Raghavan10} period distribution), 59\,per cent of systems can host a planet at 1\,au and 75\,per cent of systems can host a planet at 100\,au (the solid red circles in Fig.~\ref{stable_fracs_circumbin}). Obviously these numbers are dominated by `systems' that are single stars. If we increase the binary fraction to $f_{\rm bin} = 1.0$ but assume the same period distribution, then only 7\,per cent of systems can 
host a planet at 1\,au, rising to 44\,per cent of systems that can host a planet at 100\,au (the open red squares).

The effect of changing the binary period distribution is slightly different compared to the case of circumprimary planets. If we use the \citet{Kroupa95a} period distribution (with a binary fraction $f_{\rm bin} = 1.0$, the blue triangles in Fig.~\ref{stable_fracs_circumbin}), we see that fewer planets are stable compared to both the \citeauthor{Raghavan10} (the red open squares) and \citeauthor{Opik24} (the green diamonds) distributions, which is again due to the fact that that the Kroupa binary population is dominated by wide binaries that could not host a circumbinary 
planet at low ($< 10$\,au) separations.

In summary, depending on the assumed binary population, the fraction of systems that can host a planet at on a circumbinary P--type orbit at 1\,au ranges from $\sim$5\,per cent to 59\,per cent. This fraction increases as a function of planet semi-major axis; the fraction of systems that can host a planet at 100\,au rises to between 34\,per cent and 75\,per cent.

\section{Discussion and Conclusions}
\label{conclude}

We have determined the fraction of systems (single stars and binaries) that could host stable planetary systems, based on the well-constrained binary statistics for G-dwarfs in the Galactic field \citep{Raghavan10}. We have also considered a range of binary fractions and separation distributions, as well as different semi-major axes for planets.
When considering planetary systems orbiting single stars, or binaries with separations well in excess of the edge of the Solar System \citep[50\,au,][]{Allen01}, the fraction of systems that could host Solar System analogues is 63\,per cent. 
If we assume a higher primordial binary fraction ($f_{\rm bin} = 0.75$ versus $f_{\rm bin} = 0.46$), the proportion of Solar System analogues is greatly reduced, to  38\,per cent of systems.

However, in the more general case where planets have a range of separations, we find that the fraction of systems that can host planets on circumprimary (S--type) orbits is high; above 70\,per cent for planets with $a = 1$\,au, which decreases with increasing planet separation. Even assuming a high primordial binary fraction (e.g.\,\,unity) and different semi-major axes distributions 
the fraction of systems that can host planets at 30\,au is still $\gtrsim$40\,per cent. If most star formation leads to binary stars \citep{Goodwin05a} then planet formation beyond 30\,au (e.g.\,\,through gravitational instability) could be hindered; however, recent models of cluster evolution 
have suggested that the primordial binary fraction is unlikely to be higher than $\sim$75\,per cent \citep{Parker11c,Kaczmarek11,King12a} and according to our models 40\,per cent of systems could form a planet at 100\,au, even with this high binary fraction.

In this context it is interesting to recall two observational results:

(i) Current estimates of the occurrence rate of planets around \emph{single} solar-type stars confirm that planets are ubiquitous. Including planet candidates and not applying any mass limit, \citet{Mayor11} find a planetary rate of $75.1 \pm 7.4$\,per cent for orbital periods $P < 10$ years (corresponding to a semi-major axis of 4 -- 5 au). It seems as if the vast majority -- if not all -- solar-type stars may host at least one planet, allowing us to count all single stars as being viable planet hosts. 

(ii) Current estimates on the occurrence rate of gas giant planets at large orbital separation around solar-type stars suggest that such objects are rare \citep{Lafreniere07,Heinze10,Chauvin10}\footnote{It should be noted that the median spectral type of stars in these surveys is not early G but rather early K; however, similar results were also found for early type B, A and F stars.}. In particular, for objects $>$3 Jupiter masses the fraction of stars that have gas giant planets orbiting at separations $\approx$\,30 au is estimated to be $\lesssim$20\,per cent and even less for larger separations \citep{Lafreniere07,Chauvin10}. Our results indicate that from a stellar binary perspective more stars are ``allowed'' to host planets at these large separations and it will be interesting to see whether more sensitive surveys in the future can rule out the existence of even lower mass planets at these locations.\\

We have also estimated the fraction of G-dwarf systems in the Solar neighbourhood that could host planets on circumbinary P--type orbits, i.e.\,\,orbiting both stars in the binary system. In one sense, the respective fractions are the inverse of the S--type orbits; the assumed binary period distributions give most binaries with separations less than several 10s of au and this is reflected in the paucity of systems that can host a circumbinary planet at 1\,au. If we allow planets to form at 100\,au from the binary, then most binary systems have much lower separations than this, and between 35 and 75\,per cent of systems could host a circumbinary planet at this separation. 

Note that we have not considered whether these dynamically stable planets on both S-- and P--type orbits could be habitable -- for recent results on this topic we refer the interested reader to e.g.\,\,\citet{Eggl13,Haghighipour13,Kaltenegger13}, and references therein.

Our estimates for the fraction of systems that could likely host planetary systems could be affected by the following additional factors:

(i) When considering the statistics for binaries in the field, we note that dynamical evolution in star forming regions may affect our determination of the fraction of systems that can host stable planets in two ways. Stars which are single now may have been a member of a binary system that was broken apart 
through dynamical interactions in a dense environment \citep[e.g.][]{Kroupa95a,Parker11c}, which could have disrupted planet formation around one or both binary components \citep[e.g.][]{Fabrycky07,Malmberg07a,Parker09c}. Alternatively, a binary that is wide enough to host a stable planetary system may not have been in the past -- 
having suffered a dynamical encounter that `softened' it \citep[i.e.\,\,increased the semi-major axis,][]{Heggie75,Hills75b}. Conversely, a tight binary may have been significantly `hardened' -- a decrease in semi-major axis -- though this process would likely make any planetary system inherently unstable.

(ii) We have also assumed that every single star is able to form, or has formed a planetary system. Again, if stars are born in dense environments, then truncation of protoplanetary discs through dynamical interactions \citep[e.g.][and references therein]{Parker12a,DeJuan12} or photoevaporation \citep[e.g.][]{Armitage00,Adams04,Adams06} -- or a combination of the two \citep{Scally01} -- could limit the number of stars that are able to form planetary systems, even without considering whether a star is in a binary or not. 

(iii) Not all star forming environments are dense enough to significantly process planetary systems \citep{Bressert10}, although if star formation occurs in a hierarchical and substructured fashion then dynamical processing need not occur in dense, embedded clusters \citep{Kruijssen12b,Parker12d}. However, we can estimate that the 
maximum fraction of systems that could be affected by dynamics is represented by the difference between a primordial binary fraction of unity, and the currently observed value in the field $f_{\rm bin} = 0.46$. As we have shown that the fraction of systems that can host a planetary system depends more strongly on the binary 
fraction, rather than the orbital separation distribution, the lower limit to the number of systems that can host planets is likely to be $\sim$40\,per cent (see Fig.~\ref{stable_fracs_circumprime}).

(iv) Our analysis only considers the dynamical impact of a secondary stellar component on a generic planetary system. Higher-order multiple stellar systems are not considered. \citet{Raghavan10} find that the contribution of triples and higher order systems to multiple systems in the field is 12\,per cent (from a total fraction of 
46\,per cent for all multiple systems). Planets in triple systems could be further de-stablised by perturbations from the third star, which would likely reduce the overall fractions of stable systems given in Fig.~\ref{stable_fracs}. Also, dynamical interactions between several planets in a planetary system can lead to instabilities, significant re-arrangements of planetary orbits and even the ejection of planets from that system \citep[e.g.][]{Raymond11}. Hence, some fraction of stars that can harbour a planetary system in our analyses may not necessarily host an intrinsically stable planetary system.

Finally, we note that we have only considered planetary systems around G-dwarfs in our analysis,  whereas most stars in the Galaxy are M-dwarfs. However, the binary statistics of M-dwarfs in the field are not as robust as for G-dwarfs, although this is currently being addressed 
\citep[e.g.][]{Bergfors10,Janson12}. Preliminary results suggest that the binary fraction of M-dwarfs is lower than for G-dwarfs \citep[see also][]{Fischer92}, and the separation distribution peaks at lower values than for G-dwarfs \citep{Janson12}. The number of single stars that could host planets would therefore be higher, 
but the smaller binary separations would prevent planets being stable at wider separations.

\section*{Acknowledgments}

We thank the anonymous referee for their thorough and helpful review, which greatly improved the manuscript.

\bibliographystyle{mn2e}
\bibliography{general_ref}

\label{lastpage}

\end{document}